
\mag 1200

\overfullrule 0pt

\input amstex
\input amsppt.sty

\NoRunningHeads

\hsize 4.5in
\vsize 7in

\expandafter\ifx\csname twist.def\endcsname\relax \else\endinput\fi
\expandafter\edef\csname twist.def\endcsname{%
 \catcode`\noexpand\@=\the\catcode`\@\space}
\catcode`\@=11

\let\logo@\relax

\mathsurround 1.6pt

\def\hcor#1{\advance\hoffset by #1}
\def\vcor#1{\advance\voffset by #1}
\let\bls\baselineskip \let\dwd\displaywidth
\def\vsk#1>{\vskip#1\bls} \let\adv\advance 
\def\vv#1>{\vadjust{\vsk#1>}} \def\vvv#1>{\vadjust{\vskip#1}}
\def\vvn#1>{\vadjust{\nobreak\vsk#1>\nobreak}}
\def\vvvn#1>{\vadjust{\nobreak\vskip#1\nobreak}}
\edef\normalbls{\bls\the\bls}

\let\vp\vphantom \let\hp\hphantom \let\^\negthickspace
\let\nl=\newline \let\nt\noindent \let\cl\centerline
\def\nn#1>{\noalign{\vskip #1pt}} \def\NN#1>{\openup#1pt}
 
\let\Sum\sum \def\sum{\Sum\limits} 
\let\Prod\prod \def\prod{\Prod\limits} \let\Int\int \def\int{\Int\limits}

\let\=\m@th \def\&{.\kern.1em}

\newbox\dib@x
\def\hleft#1:#2{\setbox\dib@x\hbox{$\dsize #1\quad$}\rlap{$\dsize #2$}
 \kern-2\wd\dib@x\kern\dwd}
\def\hright#1:#2{\setbox\dib@x\hbox{$\dsize #1\quad$}\kern-\wd\dib@x
 \kern\dwd\kern-\wd\dib@x\llap{$\dsize #2$}}

\newbox\sectb@x
\def\sect#1 #2\par{\removelastskip\vskip.8\bls
 \vtop{\bf\setbox\sectb@x\hbox{#1} \parindent\wd\sectb@x
 \ifdim\parindent>0pt\adv\parindent.5em\fi\item{#1}#2\strut}%
 \nointerlineskip\nobreak\vtop{\strut}\nobreak\vskip-.6\bls\nobreak}

\def\gad#1{\global\advance#1 1}
\def\l@b@l#1@#2>{\def\n@{\csname #2no\endcsname}%
 \expandafter\ifx\csname @#1@\endcsname\relax\gad\n@
 \expandafter\xdef\csname @#1@\endcsname{\the\Sno.\the\n@}\fi}
\def\l@bel#1@#2>{\l@b@l#1@#2>\[#1]}
\def\[#1]{\csname @#1@\endcsname}
\def\dff{\expandafter\d@f} \def\daf{\expandafter\d@f}

\newcount\Sno \newcount\Lno \newcount\Fno
\def\Sect{\gad\Sno\Fno=0\Lno=0 \sect{\the\Sno.} }
\def\<#1>{\l@bel #1@F>} \def\?#1?{\l@bel #1@L>}
\def\Tag#1{\tag\<#1>} \def\Tagg#1{\tag"\rlap{(\<#1>)}"}
\def\Df#1{Definition \?#1?} \def\Th#1{Theorem \?#1?}
\def\Lm#1{Lemma \?#1?} 
 \def\Ex#1{\example{Example #1}}
\def\Proof#1.{\demo{Proof #1}} \def\qed{\hbox{}\hfill$\square$}
\def\endproof{\qed\enddemo}

\let\alb\allowbreak \def\({\allowbreak(} \def\alh{\hfil\alb\hfilneg}
 \let\alds\allowdisplaybreaks

  \let\ox\otimes
\let\sub\subset 
\let\le\leqslant 
\let\der\partial  \let\8\infty
\let\bra\langle \let\ket\rangle
 
 \let\map\mapsto 
\let\=\m@th \let\wh\widehat \def\_#1{_{\rlap{$\ssize#1$}}}

 \let\Gm\Gamma
\let\dl\delta 
\let\epe\epsilon \let\eps\varepsilon \let\epsilon\eps

 \let\phi\varphi
\let\om\omega 
 \let\rho\varrho

\def\C{\Bbb C}

\def\Z{\Bbb Z}
\def\Zn{\Z_N^n}

\def\e{\bold e}
\def\s{\bold s}
\def\t{\bold t}

\def\A{\Cal A}
\def\F{\Cal F}

\def\We{\Cal W}

\def\M{\hat M}
\def\Gh{\hat\Gm}
\def\ah{\hat a}
\def\bh{\hat b}
\def\ph{\hat p}

\def\J{\tilde J}
\def\Pti{\tilde\Phi}
\def\Q{\tilde Q}
\def\S{\widetilde S}
\def\W{\widetilde W}

\def\V{\overset\,{\sssize\circ}\to{\smash{V}\vp v}}

\def\lsym#1{#1\alb\ldots#1\alb}
\def\lc{\lsym,}   
\def\E(#1){\mathop{\hbox{\rm End}\,}(#1)} 
 
\def\vst#1{{\lower2.1pt\hbox{$\bigr|_{#1}$}}}
\def\1{^{-1}} \def\0{^{\vp1}}

\def\eq/{equation}
\def\lhs/{the left hand side} \def\rhs/{the right hand side}
\def\itw/{intertwiner}
\def\rep/{representation} \def\ir/{irreducible} \def\irp/{\ir/ \rep/}
\def\YB/{Yang-Baxter \eq/} \def\hm/{homomorphism}
\def\gb/{generated by} \def\asa/{an associative algebra with unit}
\def\wrt/{with respect to} \def\perm/{permutation}
\def\sol/{solution} \def\ism/{isomorphism} \def\isc/{isomorphic}

\def\ido/{an identity operator}
\def\CP{\C P^{2n-1}} \def\Cp{\C P^1}
\def\hc/{homogeneous coordinates}
\def\tYB/{the twisted \YB/}
\def\lca/{the lattice current algebra}
\def\DJ/{DJMM}

\csname twist.def\endcsname

\font\bbf=cmbx12

\document

\line{{\sl hep-th/9403011}\hfil\=HU$-$TFT$-$94$-$9}
\vsk2>
\cl{\bbf On Solutions to the Twisted Yang-Baxter equation}
\vsk1.5>
\cl{V\&O\&Tarasov}
\vsk>
\cl{\it Researh Institute for Theoretical Physics}
\vsk.1>
\cl{\it P.O\&Box 9 (Siltavourenpenger 20 C), {\=SF$-$00014}
\,University of Helsinki}
\vsk1.5>
\cl{\eightpoint\sl
Dedicated to L\&D\&Faddeev on his 60$^{\hbox{th}}$ birthday}
\vsk1.5>
{\narrower\eightpoint\nt
{\bf Abstract.}\enspace
Solutions to \tYB/ arising from \itw/s for cyclic \rep/s of $U_q(\wh{sl}_n)$
are described via two coupled \lca/s.
\vsk1.3>}
\footnotetext""{\vv-1>\nl
On leave of absence from
Physics Department, St\&Petersburg University, Russia\nl
\sl E-mail: tarasov\@finuhcb.bitnet}

\sect{\hskip\parindent Introduction}
\par
In this note I describe a few \sol/s to the twisted (generalized) \YB/
$$
S(q,r)\S(p,r)S(p,q)=\rho(p,q,r)\S(p,q)S(p,r)\S(q,r)\,.
$$
Recently, such an \eq/ was introduced explicitly in \cite{KS}.
It was shown there
that starting from its \sol/ one can make a \sol/ to the usual \YB/
(cf\. ``box'' construction), and thus a quantum integrable model (a solvable
model in statistical mechanics) can be defined. An example of a \sol/
for the twisted \YB/ related to cyclic \rep/s of $U_q(\wh{sl}_3)$
{\=($sl_3$ chiral Potts model)} was also given therein.
\par
On the other side it was shown in \cite{\DJ/}\,, \cite{T}
that \itw/s for cyclic \rep/s of
$U_q(\wh{sl}_n)$ for any $n$ give \sol/s to \tYB/.
In this paper I clear up
the simple algebraic base of these \sol/s.
The algebra $\A$, which comes into being,
constitutes of two coupled \lca/s \cite{FV}.
The \rep/ theory of this new algebra depends on the residue of the number
of generators modulo $3$ (instead of the residue modulo $2$ for \lca/).
\par
The paper is organized as follows. In the first section I give necessary
definitions and formulate the basic theorem (Theorem 1.5). The \rep/ theory
for the algebra $\A$ is described in the second section. The third section
contains a proof of Theorem 1.5\,.

\Sect Basic \sol/
\par
Let $\om$ be a primitive {\=$N$-th} root of unity and $q^N=(-1)^{N-1}$.
\proclaim{\Df a}
An algebra $\A$ is \asa/ \gb/ elements
$J_i,\,\J_i$, \ $i\in \Z_n$ subject to the relations
$$
\NN1>
\gather
\alignedat2
J_iJ_{i+1} &= \om J_{i+1}J_i\,, &\qquad\quad
\J_i\J_{i+1} &= \om \J_{i+1}\J_i\,,
\\
J_i\J_i &= \om\1 \J_iJ_i\,, &\qquad\quad
J_{i+1}\J_i &=\om \J_iJ_{i+1}\,,
\endalignedat
\\
J_iJ_j = J_jJ_i\,, \qquad \J_i\J_j = \J_jJ_i \qquad \text{for $i\ne j\pm 1$,}
\\
J_i\J_j = \J_jJ_i \qquad\quad \text{for $\,i\ne j,j-1\,$,}
\\
J_i^N = (-1)^{N-1}\,, \qquad\quad \J_i^N = (-1)^{N-1}\,,
\\
J_1\ldots J_n = \om\1\,, \qquad\quad \J_1\ldots \J_n = \om\1\,.
\endgather
$$
\endproclaim
Let $M=\CP$ with \hc/ $a_1\lc a_n$,$b_1\lc b_n$ and
$\M=\CP$ with \hc/ $\ah_1\lc \ah_n$,$\bh_1\lc \bh_n$. Consider a covering
$$
\aligned
\widehat{\hp1}:M &\to \M \,,
\\
p=(a_1\lc b_n) &\map \ph=(a_1^N\lc b_n^N)\,.
\endaligned
\Tag{01}
$$
Let $\Gm\sub M$ be a preimage of a projective line $\Gh=\Cp\sub\M$ for this
covering. Suppose that for any $i,k$
$$
{\der(\ah_i,\bh_i)\over\der(\ah_k,\bh_k)}\ne 0\,,\qquad
{\der(\ah_i,\bh_{i+1})\over\der(\ah_k,\bh_{k+1})}\ne 0
$$
(Jacobians are calculated on $\Gh$ and the periodicity conditions
$a_{n+i}=a_i$, $b_{n+i}=b_i$ are assumed.)
Introduce $\Phi_i,\Pti_i$, \ $i=1\lc n$ such that
$$
{\der(\ah_i,\bh_i)\over\der(\ah_k,\bh_k)} = \Phi_i^N\Phi_k^{-N}\,,\qquad
{\der(\ah_i,\bh_{i+1})\over\der(\ah_k,\bh_{k+1})} = \Pti_i^N\Pti_k^{-N}\,.
\Tag{1}
$$
\proclaim{\Df b}
Let $\s\in\Z^n$ be subject to the inequalities
$s_1\le\ldots\le s_n$, \ $s_0$\,-- an integer, such that $s_0\le s_1$,
$s_0=s_n \pmod N$ and $p,p'\in \Gm$. Set
$$
\NN2>
\align
W(p,p',\s) &= \biggl({\Phi_i^N\over{b_i^N{a'_i}^N-a_i^N{b'_i}^N}}\biggr)
^{\tsize{{s_n-s_0}\over N}}
\prod_{i=1}^n\prod_{j=1}^{s_i-s_{i-1}}
{{b_ia'_i\om-a_ib'_i\om^j}\over\Phi_i}\,,
\\
\W(p,p',\s) &= \biggl({\Pti_i^N\over{b_{i+1}^N{a'_i}^N-a_i^N{b'\_{i+1}}^N}}
\biggr)^{\tsize{{s_n-s_0}\over N}}
\prod_{i=1}^n\prod_{j=1}^{s_i-s_{i-1}}
{{b_{i+1}a'_i\om-a_ib'_{i+1}\om^j}\over\Pti_i}\,.
\endalign
$$
\endproclaim
\nt
Evidently, \rhs/ in the above formulae do not depend on a choice of $s_0$, and
the first factors therein actually
do not depend on $i$ due to equalities (\[1]).
\proclaim{\Df c}
 For any $\s\in\Z^n$ set
$$
\NN1>
\align
J(\s) &= \om^{s_1s_n}
\prod_{i=1}^n\om^{(1-s_i)s_i/2}J_1^{s_1}\ldots J_n^{s_n}\,,
\\
\J(\s) &= \om^{s_1s_n}
\prod_{i=1}^n\om^{(1-s_i)s_i/2}\J_1^{s_1}\ldots\J_n^{s_n}\,.
\endalign
$$
\endproclaim
Let $\e_1\lc \e_n$ be the canonical base in $\Z^n$. The functions $W(p,p',\s)$
and $\W(p,p',\s)$ enjoy the following relations
$$
\aligned
{W(p,p',\s+\e_i)\over W(p,p',\s)} &= \Phi_{i+1}\Phi_i\1\,
{{b_ia'_i-a_ib'_i\om^{s_i-s_{i-1}}}\over
{b_{i+1}a'_{i+1}-a_{i+1}b'_{i+1}\om^{s_{i+1}-s_i-1}}}\,,
\\ \nn2>
{\W(p,p',\s+\e_i)\over \W(p,p',\s)} &= \Pti_{i+1}\Pti_i\1\,
{{b_{i+1}a'_i-a_ib'_{i+1}\om^{s_i-s_{i-1}}}\over
{b_{i+2}a'_{i+1}-a_{i+1}b'_{i+2}\om^{s_{i+1}-s_i-1}}}
\endaligned
\Tag{02}
$$
which define these functions for all $\s\in\Z^n$. $W(p,p',\s)$,
$\W(p,p',\s)$, $J(\s)$, $\J(\s)$ are invariant under the translations
$\s\map\s+N\e_i$; further on they are considered as functions on $\Zn$.
($W(p,p',\s)$, $\W(p,p',\s)$, $J(\s)$, $\J(\s)$ are also invariant under
the translation $\s\map\s+\sum_{i=1}^n \e_i$,
but this property is not employed explicitly in the paper).
\proclaim{\Df d}
Set
$$
S(p,p')=\sum_{s\in\Zn} W(p,p',\s)J(\s)\,,\qquad
\S(p,p')=\sum_{s\in\Zn} \W(p,p',\s)\J(\s)\,.
$$
\endproclaim
\proclaim{\Th A}
The following \eq/s hold
$$
\alds
\gather
\aligned
S(p,p')S(p',p) &= N^{n+1}
\prod_{i=1}^n{{b_ia'_i-a_ib'_i}\over{b_i^N{a'_i}^N-a_i^N{b'_i}^N}}\cdot
{{\prod_{i=1}^n b_i^N{a'_i}^N - \prod_{i=1}^n a_i^N{b'_i}^N}\over
{\prod_{i=1}^n b_ia'_i - \prod_{i=1}^n a_ib'_i}}\,,
\\ \nn2>
\S(p,p')\S(p',p) &= N^{n+1}
\prod_{i=1}^n{{b_{i+1}a'_i-a_ib'_{i+1}}\over
{b_{i+1}^N{a'_i}^N-a_i^N{b'\_{i+1}}^N}}\cdot
{{\prod_{i=1}^n b_{i+1}^N{a'_i}^N - \prod_{i=1}^n a_i^N{b'\_{i+1}}^N}\over
{\prod_{i=1}^n b_{i+1}a'_i - \prod_{i=1}^n a_ib'_{i+1}}}\,,
\endaligned
\\ \nn5>
S(p',p'')\S(p,p'')S(p,p')=\rho(p,p',p'')\S(p,p')S(p,p'')\S(p',p'')
\endgather
$$
\vskip2pt \nt
where $\rho(p,p',p'')$ is a scalar factor.
\endproclaim
\nt
The proof is given in the last section.

\Sect Representations of the algebra $\A$
\par
\proclaim{\Df w}
An algebra $\We$ is \asa/ \gb/ elements $Z,X$ subject to
the relations
$$
ZX=\om XZ\,, \qquad Z^N=1\,, \qquad X^N=1\,.
$$
\endproclaim
Let $\V$ be a unique \irp/ of $\We$ and $e_1\lc e_N$ -- its natural base:
$$
Ze_i=\om^ie_i\,,\qquad\quad Xe_i=e_{i+1}
\Tag{irp}
$$
where $e_{N+1}=e_1$.
\proclaim{\Th B}
\itemitem{a)} $\A$ is \isc/ to $\We^{\ox(n-1)}$ if $n\ne 0\pmod 3$;
\vsk.15>
\itemitem{b)}$\A$ is \isc/ to $\We^{\ox(n-2)}\ox\C\,[\Z_N^2]$ \,
if $n=0\pmod 3$.
\endproclaim
\Proof.
In the first case the algebra $\A$ is obviously \gb/ elements
$J_i,\J_i$, \ $i=1\lc n-1$. In the second case it is \gb/
$J_i,\J_i$, \ $i=1\lc n-2$ and two central elements
$$
C_1=\prod_{k=1}^{n/3}J_{3k-2}\J\1_{3k-1}\,,\qquad
C_2=\prod_{k=1}^{n/3}J_{3k-1}\J\1_{3k}
$$
obeying the conditions $C_1^N=1$, $C_2^N=1$. Hence, it suffices to show that
for any $m<n$, such that $m\ne 2\pmod 3$,
the subalgebra \gb/ $J_i,\J_i$, \ $i=1\lc m$, is \isc/ to $\We^{\ox m}$.
The \ism/ can be written explicitly on generators:
$$
\NN1>
\alignat3
\qquad J_i &\map qZ_{i-1}Z_i\,, &\J_i &\map qX_i\1Z_{i+1}\,,&\qquad
i &= 1\pmod 3\,,
\\
\qquad J_i &\map qX_{i-1}X_i\1\,, &\J_i &\map qZ_{i-1}Z_{i+1}\,,& \qquad
i &= 2\pmod 3\,,
\Tagg{ism}\\
\qquad J_i &\map Z_{i-2}\1X_{i-1}Z_i\1X_i\1, \quad &
\J_i &\map qX_iX_{i+2}\1\,,& \qquad i &= 0\pmod 3
\endalignat
$$
where $i=1\lc m$ and
\vadjust{
\vbox{
\vsk-.3>
$$
\alignat3
Z_i &= 1^{\ox(i-1)}\ox{} && Z\ox 1^{\ox(m-i)}\,, &\qquad
& X_i = 1^{\ox(i-1)}\ox X\ox 1^{\ox(m-i)}\,,
\\ \nn1>
Z_0 &= 1\,, && Z_{m+1}=1\,, &\qquad & X_{m+2}= 1\,.
\endalignat
$$
\vsk-1.5>\nt\qed}}
\enddemo
To interpret $S(p,p')$ and $\S(p,p')$  in terms of integrable models we are
looking for \rep/s of the algebra $\A$ subject to the restrictions
\itemitem{a)} The \rep/ space is equal to $V^{\ox3}$
for some vector space $V$;
\itemitem{b)} Generators {\=$J_i$'s} act as \ido/ in the third
tensor factor $V$;
\itemitem{c)} Generators {\=$\J_i$'s} act as \ido/ in the first
tensor factor $V$.
\par\nt
Let us give examples of such \rep/s for the algebra $\A$.
\Ex{1}
Let $\nu:\A\to\E(V)$ be a \rep/ of the algebra $\A$.
 Fix two commutative sets $\{Q_i\in\E(V)\}_{i=1}^n$ and
$\{\Q_i\in\E(V)\}_{i=1}^n$. Then a map
$$
\aligned
J_i &\map Q_i\ox \nu(J_i)\ox 1\,,
\\ \nn1>
\J_i &\map 1\ox \nu(\J_i)\ox \Q_i
\endaligned
\Tag{04}
$$
gives a required \rep/ of $\A$. The most important case is $\nu$ being
a unique \irp/ of $\A$ ($V={\C^N}^{\ox(n-1)}$ for $n\ne0\pmod3$ and
$V={\C^N}^{\ox(n-2)}$ for $n=0\pmod3$). It also should be noted that the choice
of operators $Q_i,\Q_i$ is essential for getting concrete matrix \sol/s
to \tYB/.
\par
Consider the case $n=3$ in more details. Change slightly the construction
given above and introduce a \hm/ $\A\to\We^{\ox3}$
$$
\alignat2
J_1 &\map Z\1X \ox Z\ox 1\,, &\qquad \J_1 &\map 1\ox X\1 \ox ZX\1\,,
\\
J_2 &\map 1\ox Z\1X \ox 1\,, &\qquad \J_2 &\map 1\ox Z\ox Z\1X\,,
\Tagg{is1} \\
J_3 &\map ZX\1 \ox X\1\ox 1\,, &\qquad \J_3 &\map 1\ox Z\1X \ox 1\,.
\endalignat
$$
(Cf\. (\[ism])). For the \irp/ $\V$ of $\We$ (cf\. (\[irp]))
$S(p,p')$ and $\S(p,p')$ are represented in $\E(\V\ox\V)$ by the following
matrices
$$
\NN2>
\gather
\bra m_1,m_2|S(p,p')|n_1,n_2\ket =
\om^{(m_1-n_1)(m_2-m_1)-m_2(m_2-n_2)}\,W(p,p',\s)\,,
\\
\s = (m_1-n_1, m_2-n_2, 0)\,,
\\ \nn2>
\bra m_1,m_2|\S(p,p')|n_1,n_2\ket =
\om^{(m_2-n_2)(m_1-m_2)-m_1(m_1-n_1)}\,\W(p,p',\tilde{\s})\,,
\\
\tilde{\s} = (0, m_2-n_2, m_1-n_1)\,.
\endgather
$$
Another matrix \sol/ can be obtained from a \hm/
$$
\alignat2
J_1 &\map Z\1X \ox Z\ox 1\,, &\qquad \J_1 &\map 1\ox X\1 \ox q\1 \,,
\\
J_2 &\map 1\ox Z\1X \ox 1\,, &\qquad \J_2 &\map 1\ox Z\ox X\1Z\1 \,,
\Tagg{is2}\\
J_3 &\map ZX\1 \ox X\1\ox 1\,, &\qquad \J_3 &\map 1\ox Z\1X \ox XZq
\endalignat
$$
(only the second column differs from (\[is1])).
Now $S(p,p')$ is represented by the same matrix as before. But for $\S(p,p')$
we have
$$
\multline
\bra m_1,m_2|\S(p,p')|n_1,n_2\ket =
\\
= q^{n_1-m_1}\om^{(m_2-n_2)(n_1-n_2)-(m_1-n_1)(m_1-n_1-1)/2}\,\W(p,p',-\s)\,.
\endmultline
$$
One can easily give more matrix solutions to \tYB/ in a similar way.
\endexample
\Ex{2}
Consider a matrix $\epe$ with integer entries such that
$\epe_{ij}+\epe_{ji}=1+\dl_{ij}$.
\alh
($\dl_{ij}$ is the Kronecker symbol). Set
$\om_{ij}=\om^{\epe_{ij}}$.
\proclaim{\Df e}
An algebra $\F$ is \asa/ \gb/ elements $F_i,G_i$, \ $i\in\Z_n$ subject to the
relations
$$
\NN2>
\gather
 F_iF_j=F_jF_i\,,\qquad \om_{ij}F_iG_j=\om_{i,j+1}G_jF_i\,,\qquad
\om_{ij}G_iG_j=\om_{i+1,j+1}G_jG_i\,,
\\
 F_i^N=1\,,\qquad G_1^N=1\,,\qquad G_i=(-1)^{N-1}\quad
\ \text{for \,$i=2\lc n$}
\\
 F_1\ldots F_n\,G_n\1 \ldots G_1\1=1\,.
\endgather
$$
\endproclaim
\proclaim{\Lm{L1}}
There exist integers $m_{ij}$, \ $i,j\in\Z_n$ such that
$$
m_{i,l+1}-m_{il}-m_{l,i+1}+m_{li}=\epe_{i+1,l+1}-\epe_{il}\,.
$$
\endproclaim
\Proof.
The right hand side of the equality above is antisymmetric \wrt/ \perm/ of
$i,l$. Hence, one can find integers $n_{il}$ such that
$n_{il}-n_{li}=\epe_{i+1,l+1}-\epe_{il}$. Set $n_{ll}$ to obey
$\sum_{l=1}^nn_{il}=0$. Now $m_{ij}$ can be certainly found from
$m_{i,l+1}-m_{il}=n_{il}$.
\endproof
\proclaim{\Lm{L2}}
The map
$$
 F_i \map \prod_{l=1}^n Z_l^{\epe_{il}}\,,\qquad
G_i \map c_i X_i\1 X_{i+1}\prod_{l=1}^n Z_l^{m_{il}}
$$
extends to a \hm/ of algebras $\F \map \We^{\ox(n-1)}$. Here
$$
c_1 = \prod_{l=1}^{n-1}\om^{m_{l,l+1}-m_{l1}}c_{l+1}\1\,,\qquad
c_i = q^{m_{i,i+1}-m_{ii}+1}\quad\ \text{for \,$i=2\lc n$}\,,
$$
\vsk-.5>\nt\vv-.5>%
and
$$
\alignat2
Z_i &= 1^{\ox(i-1)}\ox Z \ox 1^{\ox(n-i-1)}\,, &\qquad
X_i &= 1^{\ox(i-1)}\ox X \ox 1^{\ox(n-i-1)}\,,
\\ \nn1>
Z_n &= Z_1\1 \ldots Z_{n-1}\1\,, &\qquad X_n &= 1\,.
\endalignat
$$
\endproclaim
\proclaim{\Lm{L3}}
The map
$$
J_i \map F_{i+1}\1G_i\ox G_i\1 F_i\ox 1\,,\qquad
\J_i \map 1\ox F_{i+1}\1G_i\ox G_i\1 F_i
$$
extends to a \hm/ of algebras $\A \map \F$.
\endproclaim
\nt
The last two Lemmas can be proved by direct calculations.
\par
Now taking a \rep/ of the algebra $\We$ in a space $V$ we get a required \rep/
of the algebra $\A$ in $V^{\ox 3}$. As in the previous example the case of
a unique \irp/ of $\We$ is the most important.
\endexample
\Ex{3}
Let us consider the special case $n=3$.
\proclaim{\Lm{L4}}
The map
$$
\alignat2
J_1 &\map XZ\ox Z\1 \ox 1\,, &\qquad \J_1 &\map 1\ox XZ\ox Z^2\,,
\\
J_2 &\map Z^{-2}\ox XZ^2q\1 \ox 1\,, &\qquad \J_2 &\map 1\ox Z \ox XZ^2q\1\,,
\\
J_3 &\map X\1 Z \ox X\1 Z\1q \ox 1\,, &\qquad
\J_3 &\map 1\ox X\1 Z^{-2} \ox X\1 Z^{-4}q\om^3
\endalignat
$$
extends to a \hm/ of algebras $\A\map \We^{\ox 3}$.
\endproclaim
Taking $\om=\exp(2\pi i/N)$, $q=-\exp(-\pi i/N)$ and the \irp/ $\V$ of $\We$
(cf\. (\[irp])) we reproduce the \sol/ to the \tYB/ described in \cite{KS}
up to notations (in particular $S\,,\S\,,\om$ in this paper correspond to
${\overline S}\,,S\,,\om\1$ in \cite{KS}\,).
\endexample

\Sect Proof of Theorem \[A]
\par
Consider again Example 2\,. Let $\V$ be the \irp/ (\[irp]) of the algebra
$\We$. Employing Theorem \[B] one can see that the \rep/ $\V^{\ox3}$ of the
algebra $\A$ is faithful. Therefore, it is enough to prove the equalities
in this \rep/. In slightly different notations it has been done in
\cite{T}.
\qed
\vsk.2>
Using explicit formulae for $S(p,p')$ and $\S(p,p')$ one can get
the following expressions for the factor $\rho(p,p',p'')$ in
Theorem \[A]\,:
$$
\NN4>
\alds
\alignat2
\qquad \rho(p,p',p'') &= N^{-n-1} &&
\sum_{\s,\t\in\Zn} {W(p,p',\s)W(p',p'',\t)\over W(p,p'',\s+\t)}\,
\om^{-\bra\s,\t\ket}\,,
\Tagg{31} \\
&= &&\ {\sum_{\s\in\Zn}W(p,p',\s)W(p',p'',-\s)\,\om^{\bra\s,\s\ket}\over
\sum_{\s\in\Zn}\W(p,p',\s)\W(p',p'',-\s)\,\om^{\bra\s,\s\ket}}\,,
\Tagg{32} \\
&= &&\ {\sum_{\s\in\Zn}W(p,p',\s)\big/W(p',p'',\s)\over
\sum_{\s\in\Zn}\W(p,p',\s)\big/\W(p',p'',\s)}\,,
\Tagg{33} \\
\qquad \rho\1(p,p',p'') &= N^{-n-1} &&
\sum_{\s,\t\in\Zn} {\W(p,p',\s)\W(p',p'',\t)\over \W(p,p'',\s+\t)}\,
\om^{-\bra\s,\t\ket}
\Tagg{34}
\endalignat
$$
where $\bra\s,\t\ket=\sum_{i=1}^n s_i(t_{i+1}-t_i)$, $t_{n+1}=t_1$.
(Cf\. also (A.23) in \cite{KS}).

\sect{\hp{4.}} Acknowledgements
\par
I greatly appreciate the hospitality of Research Institute for Theoretical
Physics.
I am also thankful to R\&M\&Kashaev for helpful discussions.
The author was supported by the Russian Foundation for Fundamental Research
and the Academy of Finland.


\newpage

\Refs
\widestnumber\key{\DJ/}

\ref
\key{\DJ/}
\by E\&Date, M\&Jimbo, K\&Miki and T\&Miwa
\paper Braid group representations arising from the generalized chiral Potts
 model
\jour Pacific J. Math. \vol 154 \yr 1992 \issue 1 \pages 37--65
\endref

\ref
\key{FV}
\by L\&D\&Faddeev and A\&V\&Volkov
\paper Abelian current algebra an the Virasoro algebra on a lattice
\jour Phys. Lett. B \vol 315 \yr 1993 \issue 3,4 \pages 311-318
\endref

\ref
\key{KS}
\by R\&M\&Kashaev and Yu\&G\&Stroganov
\paper Generalized \YB/
\jour Mod. Phys. Lett. A \vol 8 \yr 1993 \issue 24 \pages 2299--2309
\endref

\ref
\key{T}
\by V\&O\&Tarasov
\paper Cyclic monodromy matrices for $sl(n)$ trigonometric {\=$R$-matrices}
\jour Commun. Math. Phys. \vol 158 \yr 1993 \issue 3 \pages 459--483
\endref

\endRefs
\bye